\documentstyle[epsf,aps]{revtex}

\begin{document}

\setlength{\textwidth}{7.0in}
\setlength{\textheight}{8.5in}
\setlength{\evensidemargin}{0in}
\setlength{\oddsidemargin}{0in}
\setlength{\topmargin}{-0.65in}

\draft{}
\def\vpsu{$^{1}$}
\def\jlab{$^{2}$}
\def\asuaz{$^{3}$}
\def\ucla{$^{4}$}
\def\cmupa{$^{5}$}
\def\cuawdc{$^{6}$}
\def\cea {$^{7}$}
\def\cnuva{$^{8}$}
\def\connecticut{$^{9}$}
\def\duke{$^{10}$}
\def\edinburgh{$^{11}$}
\def\fiu{$^{12}$}
\def\fsu{$^{13}$}
\def\frascati{$^{14}$}
\def\genova{$^{15}$}
\def\gwudc{$^{16}$}
\def\itep{$^{17}$}
\def\jmuva{$^{18}$}
\def\knukorea{$^{19}$}
\def\umma{$^{20}$}
\def\mit{$^{21}$}
\def\unhdurham{$^{22}$}
\def\nsuva{$^{23}$}
\def\ohio{$^{24}$}
\def\oduva{$^{25}$}
\def\ipn{$^{26}$}
\def\uppa{$^{27}$}
\def\rpi{$^{28}$}
\def\rubltx{$^{29}$}
\def\urva{$^{30}$}
\def\usc{$^{31}$}
\def\utep{$^{32}$}
\def\uvch{$^{33}$}
\def\cwm{$^{34}$}
\def\yerevan{$^{35}$}


\title{{\large Exclusive electroproduction of $\phi$ mesons at 4.2 GeV}}

\author{K.~Lukashin,\vpsu$^,$\jlab\ \thanks{Present address: 
Department of Physics, Catholic 
University of America, Washington, D.C. 20064.}
        E.S.~Smith,\jlab\ 
        G.S.~Adams,\rpi\  
        E.~Anciant,\cea \
        M.~Anghinolfi,\genova\        
        B.~Asavapibhop,\umma\ 
        G.~Audit,\cea \   
        T.~Auger,\cea \ 
        H.~Avakian,\frascati\
        J.~Ball,\asuaz\
        S.~Barrow,\fsu\ 
        M.~Battaglieri,\genova\ 
        K.~Beard,\jmuva\ 
        M.~Bektasoglu,\oduva\ 
        W.~Bertozzi,\mit\ 
        N.~Bianchi,\frascati\ 
        A.~Biselli,\rpi\ 
        S.~Boiarinov,\itep\
        B.E.~Bonner,\rubltx\ 
        S.~Bouchigny,\jlab\ 
        D.~Branford,\edinburgh\
        W.J.~Briscoe,\gwudc\
        W.K.~Brooks,\jlab\ 
        V.D.~Burkert,\jlab\ 
        J.R.~Calarco,\unhdurham\  
        D.S.~Carman,\ohio\
        B.~Carnahan,\cuawdc\ 
        L.~Ciciani,\oduva\
        R.~Clark,\cmupa\
        P.L.~Cole,\utep$^,$\jlab\ 
        A.~Coleman,\cwm\ \thanks{Present address: 
Systems Planning and Analysis, 2000 North 
Beauregard Street, Suite 400, Alexandria, VA 22311.} 
        D.~Cords,\jlab\ 
        P.~Corvisiero,\genova\ 
        D.~Crabb,\uvch\ 
        H.~Crannell,\cuawdc\
        J.~Cummings,\rpi\
        P.V.~Degtiarenko,\jlab\  
        L.C.~Dennis,\fsu\ 
        E.~De\, Sanctis,\frascati\ 
        R.~DeVita,\genova\ 
        K.S.~Dhuga,\gwudc
% ...
\\*[0.2cm](The CLAS Collaboration)
}


\address{
\vpsu Virginia Polytechnic Institute and State University, 
Department of Physics, Blacksburg, VA 24061, USA\\
\jlab Thomas Jefferson National Accelerator Facility, 
Newport News, VA 23606, USA\\
\asuaz Arizona State University, Department of Physics and Astronomy, 
Tempe, AZ 85287, USA\\
\ucla University of California at Los Angeles, 
Department of Physics and Astronomy, Los Angeles, CA 90095, USA\\
\cmupa Carnegie Mellon University, Department of Physics, 
Pittsburgh, PA 15213, USA\\
\cuawdc Catholic University of America, Department of Physics, 
Washington, DC 20064, USA\\
\cea  CEA Saclay, DAPNIA-SPhN, F91191 Gif-sur-Yvette Cedex, France\\
\cnuva Christopher Newport University, Newport News, VA 23606, USA\\
\connecticut University of Connecticut, Physics Department, Storrs, 
CT 06269, USA\\
\duke Duke University, Physics Department, Durham, NC 27706, USA\\
\edinburgh Edinburgh University, Department of Physics and Astronomy, 
Edinburgh EH9 3JZ, United Kingdom\\
\fiu Florida International University, Department of Physics, Miami, 
FL 33199, USA\\
\fsu Florida State University, Department of Physics, Tallahassee, 
FL 32306, USA\\
\frascati Istituto Nazionale di Fisica Nucleare, 
Laboratori Nazionali di Frascati, C.P. 13, 00044 Frascati, Italy\\
\genova Istituto Nazionale di Fisica Nucleare, Sezione di Genova
e Dipartimento di Fisica dell'Universita, 16146 Genova, Italy\\
\gwudc The George Washington University, Department of Physics, 
Washington, DC 20052, USA\\
\itep Institute of Theoretical and Experimental Physics, Moscow, 117259, 
Russia\\
\jmuva James Madison University, Department of Physics, Harrisonburg, 
VA 22807, USA\\
\knukorea Kyungpook National University, Department of Physics, 
Taegu 702-701, South Korea\\
\umma University of Massachusetts, Department of Physics, Amherst, 
MA 01003, USA\\
\mit M.I.T.-Bates Linear Accelerator, Middleton, MA 01949, USA\\
\unhdurham University of New Hampshire, Department of Physics, 
Durham, NH 03824, USA\\
\nsuva Norfolk State University, Norfolk, VA 23504, USA\\
\ohio Ohio University, Department of Physics, Athens, OH 45701, USA\\
\oduva Old Dominion University, Department of Physics, Norfolk, 
VA 23529, USA\\
\ipn Institut de Physique Nucleaire d'Orsay, IN2P3, BP 1, 91406 Orsay, France\\
\uppa University of Pittsburgh, Department of Physics, 
Pittsburgh, PA 15260, USA\\
\rpi Rensselaer Polytechnic Institute, Department of Physics, 
Troy, NY 12181, USA\\
\rubltx Rice University, T.W. Bonner Nuclear Laboratory, 
Houston, TX 77005-1892, USA \\
\urva University of Richmond, Department of Physics, Richmond, VA 23173, USA\\
\usc University of South Carolina, Department of Physics, 
Columbia, SC 29208, USA\\
\utep University of Texas, Department of Physics, El Paso, TX 79968, USA\\
\uvch University of Virginia, Department of Physics, 
Charlottesville, VA 22903, USA\\
\cwm College of William and Mary, Department of Physics, 
Williamsburg, VA 23185, USA\\
\yerevan Yerevan Physics Institute, 375036 Yerevan, Armenia\\
}
\date{\today}
\maketitle

\begin{abstract}
\begin{center}
{\large Abstract}
\end{center}
We studied the exclusive reaction $ep \rightarrow e'p'\phi$ using the 
$\phi\rightarrow K^+K^-$ decay mode. 
The data were collected using a 4.2 GeV incident electron beam and the 
CLAS detector at Jefferson Lab. Our
experiment covers the range in $Q^{2}$ from 0.7 to 2.2 GeV$^2$, and $W$ 
from 2.0 to 2.6 GeV. 
Taken together with all previous
data, we find a consistent picture of $\phi$ production on the proton. 
Our measurement shows the 
expected decrease of the $t$-slope with the vector meson formation time 
$c\Delta \tau$ below 2 fm. At 
$\langle c \Delta \tau \rangle$ = 0.6 fm, we measure $b_{\phi}=$ 2.27 
$\pm 0.42$ GeV$^{-2}$. 
The cross section dependence on $W$ as $W^{0.2\pm0.1}$ at $Q^2 = 1.3$ GeV$^2$ 
was determined by 
comparison with $\phi$ production at HERA after correcting for threshold 
effects. 
This is the same dependence as observed in photoproduction.\\*[0.2cm]
PACS number(s): 13.60.Le, 13.60.-r, 12.40.Vv, 14.40.Cs
\end{abstract}

\twocolumn

\section{Introduction}

\begin{figure}[htb] 
\vspace{4.2cm}
\includegraphics{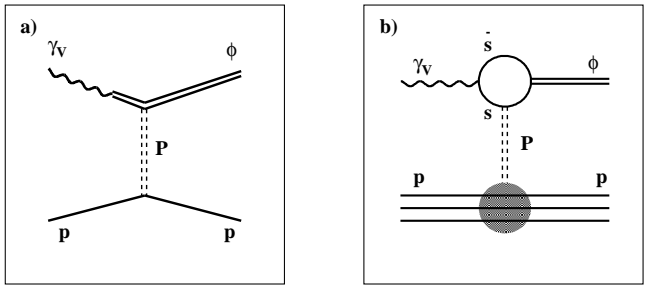}
\caption{\small{Representation of $\phi$ production by: a) the VMD model, and 
b) the Donnachie and Landshoff pomeron-exchange model.}}
\label{fig:diagrams}
\end{figure}

Vector meson photo- and electroproduction have been important tools used to 
understand the hadronic  
properties of the photon \cite{Bauer}. For low values of the four-momentum 
transfer squared,  
the photon interacts with the target predominantly through vector 
meson intermediate states that diffractively scatter off the target. This 
process, shown in 
Fig.$\:$\ref{fig:diagrams}a, was originally described within the framework 
of the Vector-Meson Dominance 
(VMD) model. The production of the $\phi$ meson through this mechanism may 
be interpreted in 
terms of the hadronic structure of the photon, that couples to a virtual 
meson with a strength 
proportional to the square of the charge of its constituent quarks.
Due to the dominant $s \bar{s}$ quark component in the $\phi$ meson, quark 
exchange
(e.g. meson-exchange) mechanisms and s-channel resonance production are 
strongly 
suppressed \cite{Freund,Barger,Joos,Kajantie}. 
As a consequence, $\phi p$ scattering at low four-momentum transfer proceeds 
primarily through 
pomeron exchange, similar to hadron-hadron diffractive scattering.

It is generally
believed that the underlying mechanism for pomeron exchange is multi-gluon 
exchange, where
the simplest possibility requires at least two gluons since all hadrons are 
color singlets.
A simplification to these calculations was introduced by Donnachie and 
Landshoff 
\cite{Donnachie1,Donnachie2},
who proposed a model whereby the pomeron couples to quarks inside the 
interacting hadrons
as shown in Fig.$\:$\ref{fig:diagrams}b.
Calculations within this context have been applied to $\phi$ electroproduction 
to study
the quark substructure of mesons \cite{Pichowsky,Laget} and to photoproduction 
at large momentum transfer
\cite{Laget2,Anciant}. In these models the cross section increases slowly with
center-of-mass energy, $W$, reflecting the pomeron trajectory. 

At high $Q^2$ the pomeron can be resolved into two-gluon exchange, and
predictions for hard diffractive electroproduction of vector mesons can be 
made within 
the context of perturbative QCD \cite{Frankfurt1}. At lower energies 
($W$ {\tiny $\stackrel{\textstyle<}{\sim}$} 10 GeV), 
quark exchange mechanisms \cite{Guidal,Kroll} become significant for the 
production of
vector mesons with valence $u$ and $d$ quarks, but play a limited role in 
the production
of $\phi$ mesons.

The hadronic structure of the photon arises from fluctuations of the virtual 
photon
into short-lived quark-antiquark  ($q \bar q$) states of mass $M_{V}$ during 
a formation time \cite{Bauer}
\begin{eqnarray}
\Delta \tau & = &  {{2\nu} \over {(Q^{2} + M_{V}^{2})}} \: ,
\label{eq:dtau}
\end{eqnarray}
where -$Q^2$ is the squared mass and $\nu$ is the laboratory-frame energy of 
the virtual photon
(see Appendix A for notation).
The effect of the formation time on the propagation of these virtual quantum 
states 
in strongly interacting media has been
observed for $\rho$ mesons propagating inside a proton \cite{Cassel} and inside
nuclear targets \cite{Ackerstaff}. To date, no clear dependence on the 
formation time
has been observed in $\phi$ meson production by virtual photons 
\cite{Cassel,Dixon,Dixon1}.

This paper presents measurements of exclusive $\phi$ meson electroproduction 
off a proton target
for $2.0 \le W \le 2.6$ GeV and $0.7 \le Q^2 \le 2.2$ GeV$^2$ where there is 
extremely limited data.
In this kinematic regime, the short formation distance 
\footnote{In the literature the formation distance is also referred to as 
coherence length.}
of the virtual $q \bar q$ state 
($0.35 \le c\Delta\tau \le 0.75$ fm) limits the time for interaction and 
probes the $\phi$
production mechanism at small formation times.

In Section \ref{sec:experiment} we present the details of our
experimental techniques and data analysis.  It concludes with values
for the measured $t$-slopes and total cross sections.  In Section 
\ref{sec:results} we
compare our results with previous data, and compare with a
geometrical model of the relation between formation
time and $t$-slope.  The model is discussed in some detail in Appendix 
\ref{sec:model}.

\section{Experiment}
\label{sec:experiment}

The experiment was performed using the CEBAF Large Acceptance Spectrometer 
(CLAS) \cite{Brooks,CLAS} in Hall B 
of the Thomas Jefferson National Accelerator Facility (Jefferson Lab). 
The data were taken with a 4.2 GeV 
electron beam incident on a 5.0 cm liquid hydrogen target in March and 
April of 1999. The CLAS torus
magnet current was set to 2250 A, bending negatively charged particles 
toward the beam axis.
The trigger required a 
single scattered electron signal, identified as a coincidence of the forward 
electromagnetic calorimeter 
(EC)\cite{EC} and Cerenkov counters \cite{CC}. Data were recorded at an 
instantaneous luminosity of 
$0.6 \cdot 10^{34}$ cm$^{-2}$ s$^{-1}$ and a typical livetime of 95\%. 
This data set has a live-time corrected integrated 
luminosity $L_{int} = 1.49 \cdot 10^{39}$ cm$^{-2}$. 

\textbf{Data Reduction.} In order to reduce the data sample to a manageable 
size, 
the data of interest were first pre-selected using very loose requirements 
on particle identification, missing mass, 
and the requirement for $W$ to be above 1.8 GeV. The $\phi$ mesons were 
identified through their
$K^+K^-$ decay mode. Because of the small acceptance of $K^{-}$ due to the 
CLAS magnetic
field setting, we required only three final-state particles to be 
detected: electron, proton, and $K^{+}$. The $K^{-}$ was 
reconstructed by identification in the $epK^{+}(X)$ missing mass. The momenta 
of charged tracks were
reconstructed from their curvature in the CLAS magnetic field using a
system of drift chambers \cite{DC}. The data reduction process 
selected about 82,000 events for further analysis. The size of this filtered 
data sample was 
compact ($\approx$ 0.5 GigaByte) and easily manageable in comparison with the 
size of 
the entire data set ($\approx$ 1 TeraByte). 

\begin{figure}[htb] 
\vspace{4.5cm}
\includegraphics{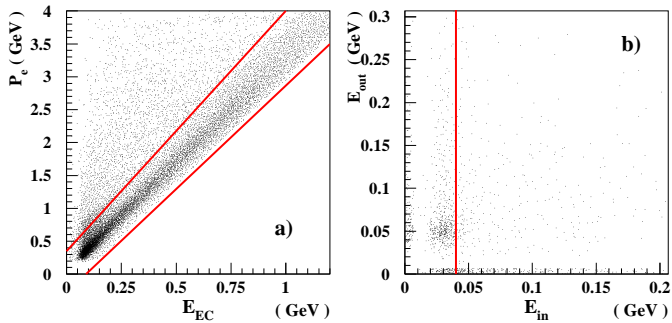}
\caption{\small{a) Electron momentum versus total deposited energy in EC. 
The solid lines show the 
applied cuts. b) Energy deposited by the TOF-identified $\pi ^{-}$'s in the 
outer EC layers versus 
energy deposited in the inner EC layers. The solid line shows the applied cut 
$E_{in} >$ 0.04 GeV, 
which retains all good electron candidates.}}
\label{fig:ec_cuts}
\end{figure}

\textbf{Electron Identification.} In addition to a fiducial requirement that 
an 
electron hit be at least 10 cm from the outer edge of the electromagnetic 
calorimeter, cuts on energy deposition 
in the EC were applied in order to avoid misidentification of $\pi^{-}$ as 
$e^{-}$. The 
total energy deposited by an electron in the EC is proportional to the 
momentum determined 
by magnetic analysis. This dependence is illustrated in 
Fig.$\:$\ref{fig:ec_cuts}a. The electron 
band with the width of the EC resolution is clearly seen. In order to cut 
out hadronic background, 
we applied cuts around this band (the solid lines in 
Fig.$\:$\ref{fig:ec_cuts}a). An additional 
improvement in e$^-$/$\pi^-$ separation was achieved by cutting out the 
$\pi^-$ signal based on the 
energy deposited in the inner layer of the calorimeter as shown in 
Fig.$\:$\ref{fig:ec_cuts}b. The 
cluster of entries to the left of the line is the $\pi^{-}$ signal in the EC. 
The solid line is the applied 
cut ($E_{in} > 0.04$ GeV) to eliminate pions. To determine this cut we used 
$\pi^{-}$ 
identified by the Time-of-Flight system (TOF) of the CLAS \cite{TOF}.

\begin{figure}[ht] 
\vspace{4.5cm}
\includegraphics{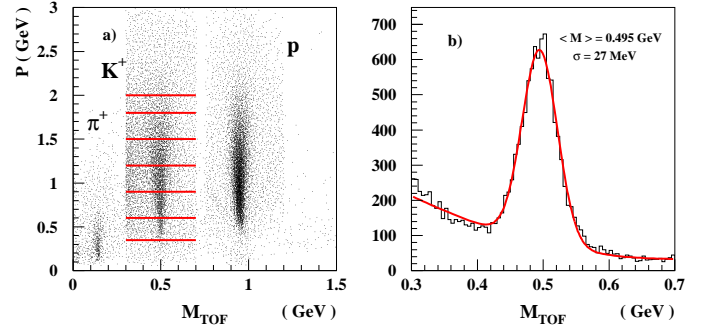}
\caption{\small{ a) Positively charged particle momentum versus reconstructed 
mass for the 
pre-selected event sample. The apparent separation between kaons, pions and 
protons at high momenta 
is due to the data pre-selection cuts. 
The horizontal lines show the binning in kaon momenta; b) $K^{+}$ 
reconstructed mass 
distribution in the momentum bin from 0.9 to 1.2 GeV. The background is due 
to pion misidentification.}}
\label{fig:PID}
\end{figure}

\textbf{Hadron Identification.} The identification of charged hadrons is 
illustrated in Fig.$\:$\ref{fig:PID}.
The distribution of positively charged particle momenta versus reconstructed 
mass is shown in 
Fig.$\:$\ref{fig:PID}a. Proton, kaon, and positive pion bands are clearly 
distinguished. The width of 
the reconstructed mass increases with momentum. However,
there is no systematic dependence after careful timing calibration of the 
detector \cite{TOF,Thesis,Simon}.

\textbf{$K^+$ Identification.} In order to optimize the signal-to-background 
ratio in kaon identification, 
the kaon momentum range was divided into six bins. In each bin the mass 
distribution was fitted to a 
Gaussian with a polynomial background to determine the characteristics of 
the $K^{+}$ peak. 
An example of this procedure is shown in Figs.$\:$\ref{fig:PID}a-b. The 
horizontal lines in 
Fig.$\:$\ref{fig:PID}a show the momentum bins for $K^{+}$ identification, 
and the fitting result for one of 
the bins is illustrated in Fig.$\:$\ref{fig:PID}b. To identify kaons, 
$\pm$2$\sigma$ cuts were applied around 
the mean value $\langle m_{K^{+}}\rangle$.

\textbf{Proton Identification.} The proton signal is very clean and does 
not have a significant background 
contribution. For proton identification we applied a simple reconstructed 
mass cut from 0.8 to 1.1 GeV.

\begin{figure}[htb] 
\vspace{8.5cm}
\includegraphics{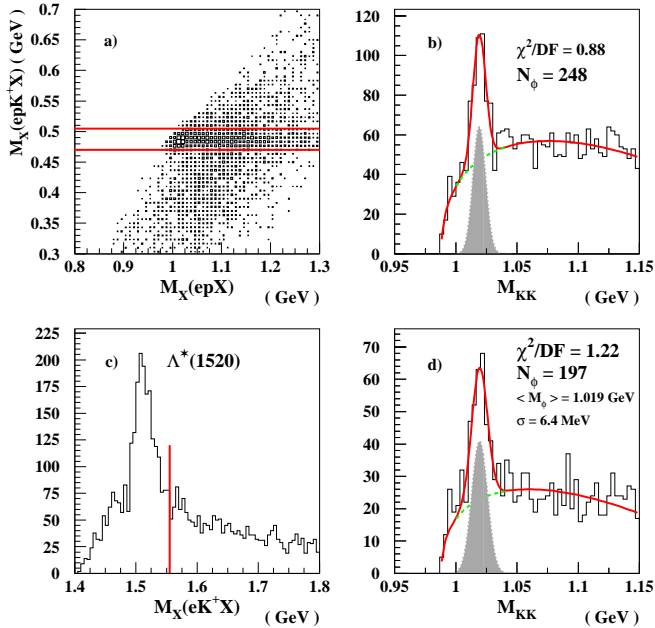}
\caption{\small{The $\phi$ channel separation technique. a) $epK^{+}X$ 
missing mass versus $epX$ missing
mass. The horizontal lines show the selection of $K^{-}$. b) $M_{KK}$ 
mass spectrum of $epK^{+}K^{-}$ events.
c) $eK^{+}X$ missing mass distribution. The line shows the $\Lambda$(1520) cut. 
d) $M_{KK}$ mass distribution with the $\Lambda$(1520) cut applied.}}
\label{fig:phi_pid}
\end{figure}

\textbf{$K^-$ Identification.} We identified the $K^{-}$ using the missing 
mass technique. The $K^{-}$ band 
is clearly seen in Fig.$\:$\ref{fig:phi_pid}a. The selection used 
$\pm 2 \sigma$ cuts around the 
$K^{-}$ peak. The invariant mass, $M_{KK}$, of the $K^+K^-$ is computed 
using
the known mass of the kaons, the measured momentum of the $K^+$ and the 
missing momentum 
of the event for the $K^-$. We note that because the masses are large 
compared to the momenta
of the particles, this quantity has significantly better resolution than 
the $epX$ missing mass.

\textbf{Identification of the Signal}. Applying the electron and hadron 
identification cuts described 
above, we identified about 3800 events of the $epK^{+}K^{-}$ final state. 
In order to eliminate events 
caused by false triggers on low energy electrons (e.g. from $\pi^{\circ}$ 
Dalitz decays) we also required 
the energy transfer, $\nu = E_{e} - E_{e'}$, to be smaller than 3.5 GeV. 
The selected sample includes 
$\phi$ mesons, high mass hyperons, and background events that come from 
particle 
misidentification.

The most important features of the final selection are shown in 
Figs.$\:$\ref{fig:phi_pid}a-c. 
In the scatter plot of $epK^{+}X$ versus $epX$ missing mass 
(see Fig.$\:$\ref{fig:phi_pid}a) the signal of the $epK^{+}K^{-}$ final 
state is clearly distinguished from 
the rest of the data. The lines solid show the $\pm 2 \sigma$ selection 
cuts in the reconstructed 
$K^{-}$ mass. Fig.$\:$\ref{fig:phi_pid}b shows the $M_{KK}$ mass 
distribution of the selected 
final state with a prominent peak due to exciting $\phi$ particles. 
To extract the total $\phi$ yield, we fitted the peak with 
a Gaussian (the integral is shown as the filled area in the plot) and 
the background with an 
empirical phase space function,
\begin{eqnarray}
f(M_{KK}) & = & A \sqrt{M^{2}_{KK} -  M_{th}^{2}} + 
B (M^{2}_{KK} - M_{th}^{2}) \: ,
\label{eq:ph_space}
\end{eqnarray} 
where the threshold $M_{th} = 0.987$ GeV. The fit gives $N_{\phi} = 248$, 
a mean value
$\langle M_{KK} \rangle = 1019.1 \pm 0.6$ MeV, and $\sigma = 6.0 \pm 0.6$ MeV, 
where the width of the 
peak is dominated by the resolution of CLAS. \footnote{The mass of the  
$\phi$ is 1019.417 $\pm$ 0.014 MeV,
and the decay width (FWHM) is 4.458 $\pm$ 0.032 MeV \cite{DataGroup}.} 
The $\phi$ signal-to-background ratio is 0.7 within $\pm 2\sigma$ from the 
mean value of the 
$\phi$ peak.  

The primary source of physical background consists of high mass hyperons, 
$ep \rightarrow e' K^+ Y^*$, with a subsequent decay 
$Y^* \rightarrow N \bar K$. The production and decay amplitudes of 
these particles are not well known. The main channel is the 
$\Lambda(1520)$ with a cross section larger than $\phi$ production. 
Additional contributions 
come from $\Lambda(1600)$, $\Lambda(1800)$, $\Lambda(1820)$, $\Sigma(1660)$, 
and $\Sigma(1750)$, which have large branching ratios for decay into the
$N \bar K$ channel \cite{DataGroup}. These backgrounds were investigated by 
Monte Carlo using
exactly the same algorithms as the experimental data in order to optimize 
selection cuts.
In order to minimize the number of $\Lambda$(1520) in the data sample, 
we require M$_{X}(eK^{+}X)$ to be greater than 1.56 GeV. The cut is shown 
for the data sample with
the solid line in Fig.$\:$\ref{fig:phi_pid}c. 

\begin{figure}[htb] 
\vspace{4.5cm}
\includegraphics{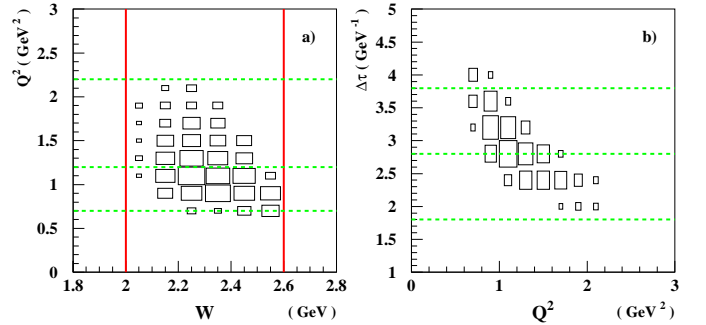}
\caption{\small{Kinematic distributions of the selected $\phi$ events: a) 
$Q^{2}$ versus $W$. b) $\Delta \tau$ versus $Q^{2}$. The dashed lines 
indicate
the binning, used later, in $Q^{2}$ and $\Delta \tau$; the solid lines show 
the range of $W$ used
in the analysis.  }}
\label{fig:kinematics}
\end{figure}

The $M_{KK}$ mass distribution with the $\Lambda$(1520) cut applied is 
shown in Fig.$\:$\ref{fig:phi_pid}d. 
The simultaneous fit of the $\phi$ peak and the background gives 
$N_{\phi} = 197$, a mean value 
$\langle M_{KK} \rangle = 1019.4 \pm 0.9$ MeV, and  $\sigma = 6.4 \pm 1.1$ MeV.
The $\phi$ signal-to-background ratio is improved, and equals 1.3 within 
$\pm 2\sigma$ of the $\phi$ peak. 
The remaining background, consistent with phase space, is due to high-mass 
hyperon states, 
non-resonant $K^+K^-$ production and experimental 
misidentification of a $\pi^{+}$ as a $K^{+}$ (events under the $K^{+}$ 
peak in Fig.$\:$\ref{fig:PID}b). 
We note that the level of the background under the $\phi$ peak depends on 
the fitting procedure, 
and will be addressed when we discuss systematic errors. 

The kinematic range of the data sample is shown in 
Fig.$\:$\ref{fig:kinematics}. 
The range of $Q^{2}$ varies from 0.7 to 2.2 GeV$^2$, $W$ from 2.0 to 2.6 GeV, 
and $\Delta \tau$ 
from 1.8 to 4.0 GeV$^{-1}$ ($c\Delta \tau$ from 0.35 to 0.79 fm). The small 
values of $c \Delta \tau$ 
indicate that the formation distance in our kinematic regime is below the 
hadron size, 
$2\, r_{h} \approx 2$ fm. 
The data binning to calculate the exponential $t$-slope (see below)
is indicated in Fig.$\:$\ref{fig:kinematics} 
by horizontal dashed lines, which show the ranges of $Q^{2}$ (integrated 
over $\Delta \tau$) 
and $\Delta \tau$ (integrated over $Q^{2}$). 
In both cases the data range in $W$ is the same 
(solid lines in Fig.$\:$\ref{fig:kinematics}a). We note that finer binning 
in $Q^{2}$ and $W$ is used
for the evaluation of the cross sections integrated over $t'$.

\begin{figure}[ht] 
\vspace{8.5cm}
\includegraphics{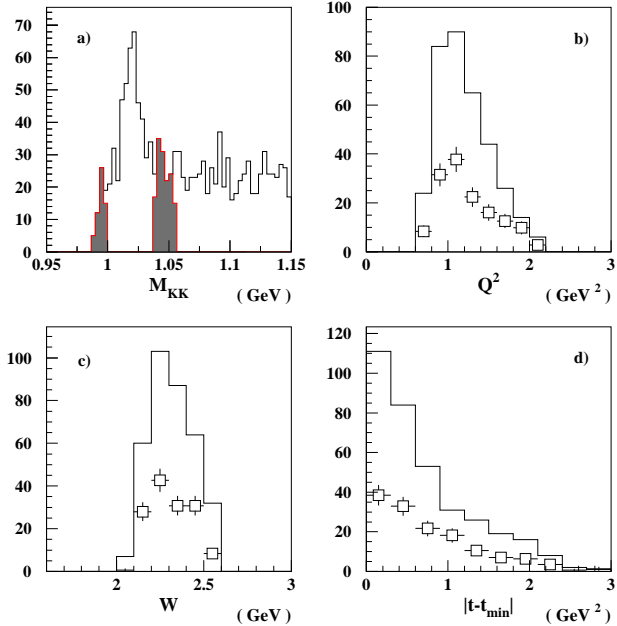}
\caption{\small{Side-band background subtraction technique.
a) Location of the side bands; b), c) and d) distributions of events
in the signal region (histograms) 
and in the side-bands (open squares) versus $Q^{2}$, $W$ and 
$\mid t - t_{min} \mid$. }}
\label{fig:background}
\end{figure}

Ideally, with enough statistics and an understanding of the background shape, 
fits would be used
to extract the signal yield in
every kinematic bin of interest. With limited statistics this is not possible, 
and we proceeded by using a side-band subtraction technique. 

\textbf{Background Subtraction.}   The side-band technique, as illustrated in 
Fig.$\:$\ref{fig:background}, was used to determine the background distribution
as a function of $Q^{2}$, $W$ and $-t'$. The signal region was determined 
within a
$\pm 2 \sigma$ cut around $\langle M_{\phi} \rangle$ after excluding 
the $\Lambda$(1520) from the final state data sample.  
The side bands were located $\pm 3 \sigma$ away from the 
$\phi$ peak, and the number within the band was scaled to the background as 
determined by the fit 
(see Fig.$\:$\ref{fig:phi_pid}d). The normalized side-band events were then 
subtracted in
each distribution of interest. This procedure is illustrated in 
Fig.$\:$\ref{fig:background} for
the entire data set and was repeated for each kinematic region defined in 
Table 1. 

\begin{center}
\begin{tabular}{| l | c | c | c | c |}
\hline \hline
Kinematic     & $Q^{2}$ and $c\Delta\tau$  & $\langle Q^{2} \rangle$  
& $\langle c\Delta\tau \rangle$ & 
$b_{\phi}$     \\ 
 region        &  range       & (GeV$^2$)  &  (fm)  &      (GeV$^{-2}$)  \\ 
\hline \hline
All Data           & 0.7 -- 2.2 GeV$^2$  &  1.02  &            &   
2.27$\pm$0.42 \\
                   & 0.35 -- 0.75 fm       &        &    0.6       
&               \\ \hline  
Low $Q^{2}$        & 0.7 -- 1.2 GeV$^2$  &  0.87  &    --       &   
2.31$\pm$0.59  \\ \hline
High $Q^{2}$       & 1.2 -- 2.2 GeV$^2$  &  1.47  &    --       &   
2.10$\pm$0.52  \\ \hline   
Low $c\Delta\tau$  & 0.35 -- 0.55 fm     &  --    &    0.49     &   
2.04$\pm$0.42  \\ \hline 
High $c\Delta\tau$ & 0.55 -- 0.75 fm     &  --    &    0.63     &   
2.12$\pm$0.46  \\ \hline \hline 
\end{tabular}
\end{center}
\vspace{1 mm}
TABLE 1. {\small The measured values of the $t$-slope parameter, $b_{\phi}$, 
fitted to the data for $-t' <$ 1.2 GeV$^2$. The errors are statistical only.}

\begin{figure}[ht] 
\vspace{4.4cm}
\includegraphics{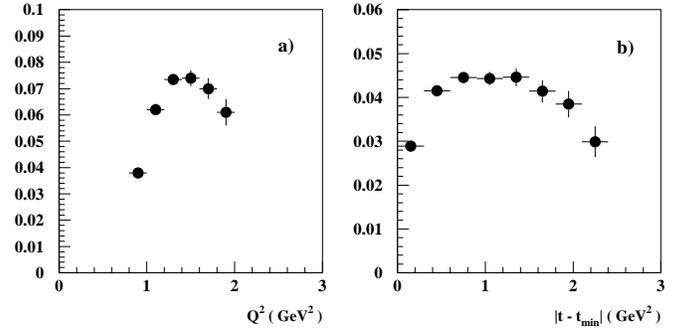}
\caption{\small{Acceptance as a function of $Q^{2}$ and $-t'$.}}
\label{fig:acc}
\end{figure}

\textbf{Acceptance.} For the calculation of the acceptance, we used a 
GEANT-based simulation 
of CLAS, taking into account trigger efficiency, problematic hardware 
channels, and the CLAS resolution. The Monte Carlo event sample was 
generated assuming the VMD model 
for $\phi$ electroproduction. Two iterations in the acceptance calculation 
were made to adjust the 
VMD parameters to be close to the data. In each kinematic region, the 
acceptance was calculated from 
the ratio of reconstructed to generated $\phi$ events with the same 
kinematics and particle 
identification cuts that were applied to the data. Fig.$\:$\ref{fig:acc} 
shows the acceptance
as a function of $Q^{2}$ and $-t'$ for the entire data set. This procedure 
was also used to calculate the acceptance as a function of $W$ and $\Delta 
\tau$ in each 
kinematic bin.

\begin{figure}[htb] 
\vspace{4.5cm}
\includegraphics{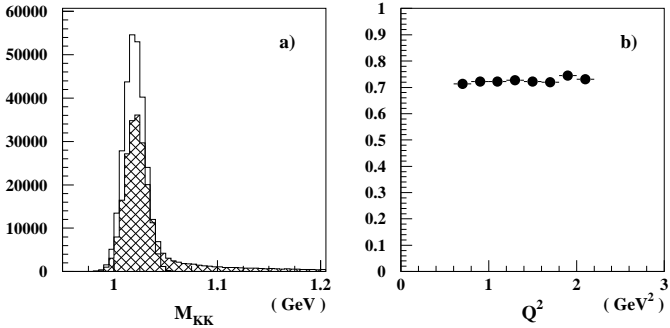}
\caption{\small{Simulated data: a) $\phi$ peak, convoluted with the 
measured CLAS resolution, 
with radiative effects turned off (solid), and turned on (hatched histogram). 
b) Inverse radiative correction factor, 1/$F^{rad}$, as a function of 
Q$^{2}$.}}
\label{fig:rad_corr}
\end{figure}

\textbf{Radiative Corrections.} For the calculation of the radiative 
corrections, we used the 
peaking approximation \cite{Ent}. We define the radiative corrections 
in each bin of 
every kinematic variable as the ratio $F^{rad} = N_{norad} / N_{rad}$, 
where $N_{rad}$ and $N_{norad}$ 
are the generated $\phi$ yields with radiative effects turned on and off, 
respectively. 
The model for the $\phi$ production cross section employed for the 
computation of 
acceptance was also used for the studies of radiative corrections.
The ratios were calculated with the same kinematics and particle 
identification cuts that were 
applied to the data. The simulated $\phi$ mass distributions with and 
without radiative effects 
are shown in Fig.$\:$\ref{fig:rad_corr}a. The inverse radiative correction 
factor, 1/$F^{rad}$,
as a function of $Q^{2}$ is shown in Fig.$\:$\ref{fig:rad_corr}b. The 
correction factors as a 
function of $-t'$ in all four kinematic regions are of the order of 1.4 
and uniform over the kinematics 
considered here.

\begin{center}
\begin{tabular}{| c | c | c | c | c |}
\hline \hline
  $Q^{2}$ bin &  $\langle W \rangle$ &  $\langle \epsilon \rangle$ 
& $\Gamma$($Q^{2}$,$W$) & $\sigma(Q^{2},W)$  \\ 
   (GeV$^2$)     & (GeV)    &              & ($10^{-4}$ GeV$^{-3}$)   
&        (nb)      \\ \hline \hline
   0.8 -- 1.0    &  2.37    &    0.51      &    1.50 $\pm$ 0.15       
&  27.6 $\pm$ 6.1  \\ \hline
   1.0 -- 1.2    &  2.31    &    0.50      &    1.12 $\pm$ 0.10       
&  24.2 $\pm$ 5.4  \\ \hline   
   1.2 -- 1.4    &  2.28    &    0.49      &   0.879 $\pm$ 0.067      
&  23.0 $\pm$ 5.2  \\ \hline 
   1.4 -- 1.6    &  2.28    &    0.44      &   0.701 $\pm$ 0.050      
&  20.8 $\pm$ 5.7  \\ \hline  
   1.6 -- 1.8    &  2.25    &    0.42      &   0.562 $\pm$ 0.033      
&  14.5 $\pm$ 6.4  \\ \hline \hline
\end{tabular}
\end{center}
\vspace{1 mm}
TABLE 2. {\small The averaged values of $W$, $\epsilon$, 
$\Gamma$($Q^{2}$,$W$), and $\sigma (Q^{2},W)$ as 
a function of $Q^{2}$. The numbers given for the virtual photon flux, 
$\Gamma$($Q^{2}$,$W$), computed 
event-by-event, are the mean and the standard deviation for the bin.}
\vspace{2 mm}

\textbf{Data Normalization.} The final step in the analysis procedure 
was the normalization of the 
$\phi$ yield to the integrated luminosity, the virtual photon flux, and 
all calculated corrections as:
\begin{eqnarray}
{ {\sigma (Q^{2},W)} }  & = & { { {N_{\phi}/BR_{\phi\rightarrow K^+ K^-}} 
\over 
{ \Delta {Q^{2}} \Delta {W} } }  {{F^{acc} F^{rad} F^{win} } 
\over {2\pi \: \Gamma(Q^{2},W) \: L_{int}}} } \: , 
\label{eq:normalization}
\end{eqnarray}
where $\Delta {Q^{2}}$ and $\Delta {W}$ are the bin widths in $Q^{2}$ and $W$,
$\Gamma(Q^{2}, W)$ is the virtual photon flux, $L_{int}$ is the integrated 
luminosity, $N_{\phi}$ 
is the $\phi$ yield in the bin, $F^{acc}$ is the acceptance factor in a 
given bin, $F^{win}$ is 
a small correction factor for production from the target windows 
($\approx 1 \%$), $F^{rad}$ 
is the radiative correction factor, and $BR = 0.492 \pm 0.007$ is the 
decay branching ratio for 
$\phi \rightarrow K^{+}K^{-}$ \cite{DataGroup}. The virtual photon flux 
was calculated on an 
event-by-event basis and averaged for each kinematic bin as
\begin{eqnarray}
\Gamma(Q^{2},W) & = & {\alpha \over {8 \pi ^{2}}} \, 
{W \over {M_{p}E_{e}^{2}}} \, 
{{W^{2} - M_{p}^{2}} \over {M_{p}Q^{2}}} \, {1 \over {1 - \epsilon}} \: ,
\label{eq:define_flux}
\end{eqnarray}
where $M_{p}$ is the mass of the proton, $E_{e}$ is the electron beam 
energy, and $\epsilon$ is 
the polarization of the virtual photon:
\begin{eqnarray}
\epsilon & = & { {4 E_e (E_e - \nu) - Q^2} \over {4 E_e(E_e - \nu) 
+ 2 \nu^2 + Q^2}}  \; .
\end{eqnarray} 

\textbf{Cross Section, $\sigma(Q^{2},W)$.} The cross section
integrated over all $t'$, $\sigma(Q^{2},W)$, was extracted in five 
bins over a $Q^{2}$ range from 0.8 to 1.8 GeV$^2$ with a bin width of 
0.2 GeV$^2$. The range in
$W$ was determined as the allowed kinematic range for each $Q^{2}$. 
The binning, values of the virtual 
photon flux used during normalization, $\Gamma$($Q^{2}$,$W$), and the 
measured cross section are given 
in Table 2. The table shows statistical errors only.

\begin{figure}[htb] 
\vspace{7.1cm}
\includegraphics{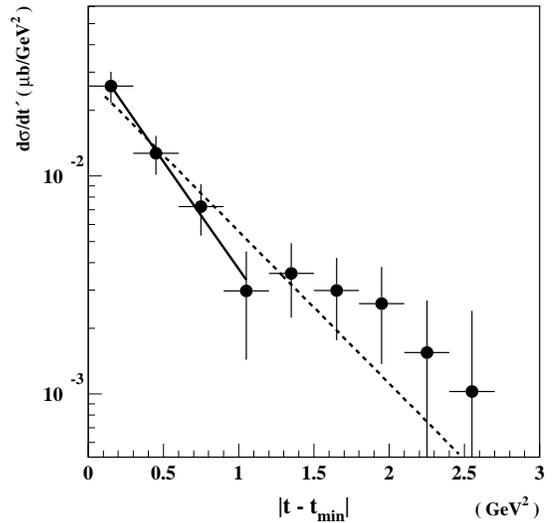}
\caption{\small{The $d\sigma / dt'$ differential cross section for 
exclusive $\phi$ electroproduction
off the proton with fits to the entire $-t'$ range (dashed) and
$-t'$ less than 1.2 GeV$^2$ (solid).}}
\label{fig:t_allq2}
\end{figure}

\textbf{Differential Cross Section, $d\sigma/dt'$.} 
The measured cross section,
$d\sigma / dt'$, is generally parameterized at small $-t'$ by 
\begin{eqnarray}
{d\sigma} \over {dt'} & = & A_{\phi} \; e^{b_{\phi}t'} \: .
\label{eq:tslope}
\end{eqnarray}
The entire $t'$ range ($0 \le -t' \le 2.6$  GeV$^{2}$) can be fit to 
a single exponential with a slope $b_{\phi}$ = 1.61$\pm$0.31 GeV$^{-2}$ 
and a $\chi^2$ = 0.9/DF. However, Eq.$\:$\ref{eq:tslope} is only expected to
be valid at small $-t'$, so we have restricted our analysis to 
$-t'$ less than 1.2 GeV$^2$ which also allows direct comparison to previous 
measurements. For this restricted range, we obtain 
$b_{\phi}$ = 2.27$\pm$0.42 GeV$^{-2}$ (solid line in Fig.$\:$\ref{fig:t_allq2}). 
We also performed fits in the four overlapping kinematic regions
specified in Table 1: two ranges in 
$Q^2$ (integrated over $c \Delta \tau$), and two ranges in $c \Delta \tau$ 
(integrated over $Q^2$). The results of these fits are given in Table 1.

We note that at larger $-t'$, there is an apparent change in the slope 
of the distribution with a 
break at $-t' \approx 1.3$ GeV$^2$. This suggests that additional 
mechanisms may be present at $-t' \ge 1$ GeV$^2$. Despite the fact that
the break is not statistically
significant, we discuss possible mechanisms for a slope change. 
A similar pattern is observed in hadron-hadron 
elastic scattering \cite{Goulianos,Castaldi}, 
where a dip is observed at $-t \approx 1.4$ GeV$^2$ followed by a 
second maximum at $-t \approx 1.8$ GeV$^2$. However, $\phi$ photoproduction 
data do not show a change in the slope for $-t \le 2$ GeV$^2$ \cite{Anciant}. 
$S$-channel production of resonances
results in a large measured value of $-t'$. However, there are no known 
resonances that decay
into $\phi N$. Finally, imperfect background subtraction could also lead 
to an enhancement at large $-t'$,
but should be subsumed into our quoted systematic errors. 

\vspace{1 mm}
\begin{center}
\begin{tabular}{| l | c | c |}
\hline \hline
Source            &  $\Delta \sigma (\%)$  &  $\Delta b_{\phi} (\%)$  \\ 
\hline \hline
Target stability  &        0.7             &    --                    \\ \hline
Target walls      &        1.0             &    --                    \\ \hline 
Acceptance        &        7.8             &    5.0                   \\ \hline 
Radiative corrections &    4.7             &    --                    \\ \hline 
Background subtraction &   5.4             &    4.6                  \\ 
\hline \hline 
Total             &       10.7             &    6.8                  \\ 
\hline \hline   
\end{tabular}
\end{center}
\vspace{1 mm}
TABLE 3. {\small Summary of the contributions to the systematic errors.}
\vspace{2 mm}

\textbf{Systematic Errors.} Estimates of our systematic errors for the cross 
section, 
$\Delta \sigma$, and the $t$-slope parameter, $\Delta b_{\phi}$, are given 
in Table 3.
The errors are averaged over the kinematics of the experiment, although the 
lowest Q$^2$ cross section
point may have about twice this systematic uncertainty due to the steepness 
of the
acceptance function (see Fig.$\:$\ref{fig:acc}a).
To estimate the systematic errors due to background subtraction, a complete 
analysis of the 
cross section and $t$-slope parameter was performed using two different 
assumptions for the 
shape of the background: phase space and a constant. The difference between 
these results is 
quoted as the systematic error due to background subtraction. The systematic 
errors due to 
acceptance and radiative corrections are discussed in Ref. \cite{Auger} and 
\cite{Thompson}, 
respectively. Additional details can be found in Ref.$\:$\cite{Thesis}. 
We note that the overall uncertainty is dominated by statistical errors.

\begin{figure}[htb] 
\vspace{7.8cm}
\includegraphics{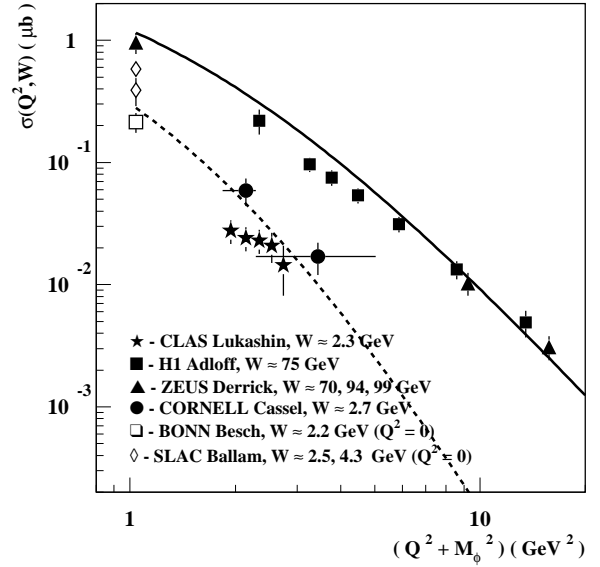}
\caption{\small{The $\phi$ meson cross section dependence on $Q^{2}$ for 
photo- and electroproduction.
Electroproduction data H1 Adloff are from Ref. [34], ZEUS Derrick from 
Ref. [35,36], and CORNELL Cassel 
from Ref. [15]. Photoproduction data BONN Besch are from Ref. [37], and 
SLAC Ballam from Ref. [38].
The solid and dashed curves are the pomeron-exchange model predictions 
for $W$=70 GeV and for $2.0 < W < 2.6$ GeV,
respectively [10].}}
\label{fig:q2_all}
\end{figure}

\section{Results}
\label{sec:results}

\begin{figure}[htb] 
\vspace{7.8cm}
\includegraphics{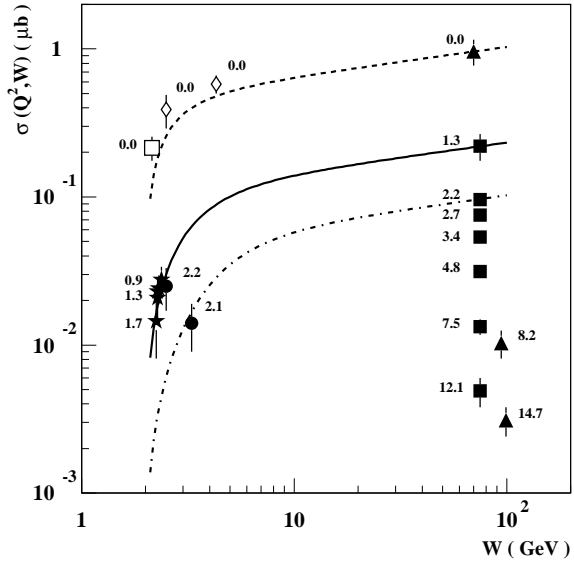}
\caption{\small{The $\phi$ meson cross section dependence on $W$ for photo- 
and electroproduction.
The $Q^2$ values of the measurements are printed near the corresponding data 
points. All data points 
are from the same References as in Fig.$\:$10. The curves, described in the 
text, correspond to $Q^2$ 
of 0, 1.3 and 2.2 GeV$^2$.}}
\label{fig:w_all}
\end{figure}

\textbf{Cross Section Dependence on $Q^{2}$ and $W$.} The world data on
elastic virtual photon production of $\phi$ mesons are shown as a function 
of
$Q^2$ in Fig.$\:$\ref{fig:q2_all}, and as a function of $W$ in 
Fig.$\:$\ref{fig:w_all}.
Selected photoproduction data are also plotted for completeness. 
\footnote{Additional data of 
$\phi$ production on nuclear targets \cite{Arneodo} are available at 
$\langle W \rangle \approx$ 14 GeV.} 
We show the data on both plots with common symbols.

All HERA data \cite{H1,Q2_ZEUS_elec,Q2_ZEUS_photo} correspond to $W$ 
ranging from 40 to 130 GeV, where the gluonic density 
in the proton at low $x = Q^2/2M_p \nu$ plays a significant role. 
Only the Cornell measurement \cite{Cassel} 
exists at low $W$, corresponding to $x$ in the valence region.
\footnote{We note that data points from Ref. \cite{Cassel} have different 
integration
ranges for the cross section as a function of  $Q^2$ and $W$ presented in 
Fig.$\:$\ref{fig:q2_all} and Fig.$\:$\ref{fig:w_all}.}
For the high-energy data, the $Q^2$ behavior of the cross section is well
described by the vector meson propagator squared. The data are not yet in 
the asymptotic perturbative QCD regime 
where the longitudinal cross section for vector meson production is dominant, 
and
should scale as $Q^{-6}$ \cite{Brodsky}. Nevertheless, the longitudinal 
contribution becomes 
increasingly important and must be treated systematically. For example, 
$\rho$ mesons in muoproduction 
at large $Q^2$ are found to be dominantly in the helicity zero spin state 
\cite{EMC}.

Pomeron-exchange models, such as those described in the introduction, 
reproduce the $Q^{-4}$ behavior of the data at large $Q^2$. The predictions 
of a model 
\cite{Laget2}, based on the Donnachie-Landshoff pomeron exchange 
(Fig.$\:$\ref{fig:diagrams}b), 
are shown in Fig.$\:$\ref{fig:q2_all} for the $W$ range of our experiment 
($2.0 < W < 2.6$ GeV) and at
$W$=70 GeV. The model describes the data
reasonably well at high $W$, and reproduces the trend at low $W$, but 
overestimates 
the new cross section results presented here. We note that our data are 
close 
to the $\phi$ production threshold, where the 
cross section increases rapidly as a function of center-of-mass energy. 
In the model of Pichowsky \cite{Pichowsky}, 
the transition from a cross section that slowly decreases with $Q^{2}$ to 
one that 
falls off as $Q^{-4}$ occurs at a threshold that increases with the 
current-quark mass of the vector meson. No clear threshold is visible
in the $\phi$ data, but the scarcity of points precludes drawing conclusions.

The photoproduction cross section increases slowly with $W$, reflecting the 
pomeron trajectory.
At higher $Q^2$, a stronger dependence on $W$ has been observed in preliminary 
analysis of
HERA data \cite{Naroska}. If the cross section is parameterized as 
$W^{\delta}$, 
$\delta$ varies from about 0.2 for photoproduction to $\delta \sim 0.7$ at a 
$Q^2$ of 8 GeV$^2$.
This increased dependence of the cross section on $W$  
has been interpreted as being due to the rise of the gluon momentum 
density in the proton at 
small $x$ \cite{Brodsky}.  

To be able to extract the $W$ dependence by comparing our measurement at 
$Q^2$ = 1.3 GeV$^2$ to
HERA data at the same $Q^2$ and $\langle W \rangle = 75$ GeV, threshold 
effects must be taken into 
account. For example, threshold behavior can be clearly seen in the 
photoproduction data \cite{Behrend} 
(see Fig.$\:$\ref{fig:w_all}). The reduced phase space near threshold 
behaves as 
$(\vec p_{\phi} / \vec q )^2$,
where $\vec p_{\phi}$ and $\vec q$ are the center-of-mass three-momenta of 
the $\phi$ and
virtual photon, respectively. This dependence of the cross section on $W$ 
can be parameterized as 
\begin{eqnarray}
\sigma (W)  & = & \sigma_{\circ} \left({ {\vec p_{\phi}} \over {\vec q}} 
\right)^{2} 
\left({{W} \over {W_{\circ}}} \right)^{\delta} \: .
\end{eqnarray} 
Correcting for the threshold factor, our measurement of 
the  cross section becomes $\sigma_{corr} (Q^2=1.3)$ = 110$\pm$27 nb, and 
using the HERA 
measurement, $\sigma (Q^2=1.3)$ = 220$\pm$51~nb \cite{H1}, we obtain 
$\delta$ = 0.2$\pm$0.1. The
quoted uncertainties were obtained by summing the statistical and 
systematic errors
in quadrature. This slope
is consistent with that measured in photoproduction. The curves of 
$\sigma (W)$ are  shown in 
Fig.$\:$\ref{fig:w_all} for $Q^{2}$ of 0, 1.3 and 2.2 GeV$^2$, and 
$\delta = 0.2$. The curves are 
normalized to the HERA data ($\sigma_{\circ}$, $W_{\circ}$) that are far 
from the production threshold.

\begin{figure}[htb] 
\vspace{6.5cm}
\includegraphics{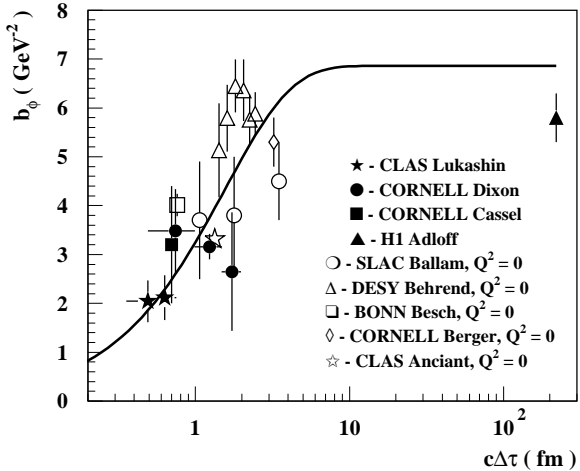}
\caption{\small{The dependence of the $t$-slope, $b_{\phi}$, on $c 
\Delta \tau$. The electroproduction data 
CORNELL Dixon are from Ref. [17,18], CORNELL Cassel from Ref. [15], 
and H1 Adloff from Ref. [34].   
The photoproduction data BONN Besch are from Ref. [37], SLAC Ballam 
from Ref. [38], DESY Behrend from
Ref. [42], and CLAS Anciant from Ref. [11].}}
\label{fig:b_dt}
\end{figure}

\textbf{Dependence of $t$-slope on $c \Delta \tau$.} The dependence of 
the $t$-slope, $b_{\phi}$, 
on formation distance, $c\Delta \tau$, for $\phi$ meson production is shown 
in Fig.$\:$\ref{fig:b_dt} 
together with previous data.
In the terminology of the uncertainty principle, $\Delta \tau$ is the time 
during which 
the virtual photon, with mass $\sqrt{Q^{2}}$, can fluctuate into a $\phi$ 
meson \cite{Bauer}. 
We expect that 
$b_{\phi}$ should decrease at low $\Delta \tau$ as the interaction 
becomes more point-like. The previous electroproduction measurements 
\cite{Cassel,Dixon,Dixon1}
do not show the expected behavior. However, a consistent picture 
emerges when we include photoproduction data as well. Both of our data 
points (solid stars) 
lie in the region of $c\Delta \tau$ below 1 fm, and show a decrease of 
$b_{\phi}$ with 
decreasing formation time when combined with other data. 
This is consistent with the well-measured dependence for $\rho$ 
meson production \cite{Cassel} as discussed in Appendix B. 
To fit the $\phi$ meson data to Eq.$\:$\ref{eq:int_size}, we
constrain the parameter $r_{h}$ to the value extracted from the
fit to the $\rho$ data (Eq.$\:$\ref{eq:rho_taufit}). This yields 
\begin{eqnarray} 
b_{\phi}(c \Delta \tau) = (6.87 \pm 0.17)
\left[ \; 1 - e^{-c\Delta \tau / 2(0.78)} \; \right]
\label{eq:phi_taufit}
\end{eqnarray}
with $\chi^{2}/DF=4.8$. The fit to the $\phi$ data is shown in 
Fig.$\:$\ref{fig:b_dt} with the solid 
curve. The ratio of $b_{\phi} / b_{\rho}$ indicates that the $\phi$ meson 
interaction size, $R^{int}_{\phi}$,
is smaller than that for the $\rho$ meson:  
\begin{eqnarray} 
{{b_{\phi}} \over {b_{\rho}}} = \left( {{R^{int}_{\phi}} \over 
{R^{int}_{\rho}}} \right)^{2}  
= 0.87 \pm 0.08 \: .
\label{eq:r_ratio}
\end{eqnarray}

\begin{figure}[htb] 
\vspace{6.5cm}
\includegraphics{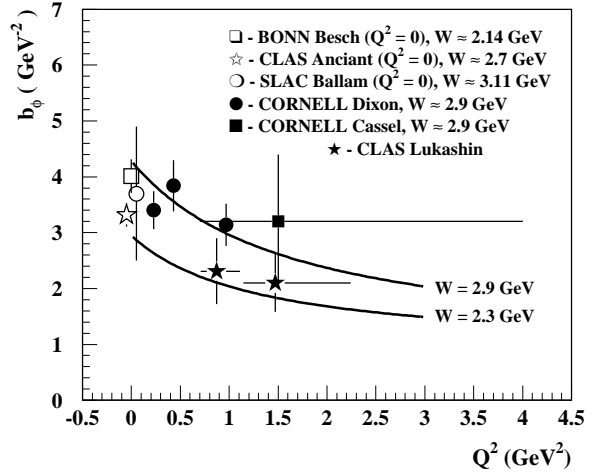}
\caption{\small{The dependence of the $t$-slope, $b_{\phi}$, on $Q^{2}$. 
Photoproduction data BONN Besch
are from Ref. [37], SLAC Ballam from Ref. [38], and CLAS Anciant from 
Ref. [11]. Electroproduction data
CORNELL Dixon are from Ref. [17,18] and CORNELL Cassel from Ref. [15].}}
\label{fig:b_q2}
\end{figure}

A summary of the existing measurements of $b_{\phi}$ together with our 
results is shown in Fig.$\:$\ref{fig:b_q2}. Previous $\phi$ electroproduction 
measurements are consistent 
with no $Q^{2}$ or $c \Delta \tau$ dependence \cite{Cassel,Dixon1}. We 
observe a low value of 
$b_{\phi} \approx$ 2.2 GeV$^{-2}$, which, taken together with the values 
measured in photoproduction, 
shows a significant dependence on $Q^{2}$. However,  
the $Q^{2}$ dependence of $b_{\phi}$ can be explained by the implicit 
dependence of $c \Delta \tau$ on 
$Q^{2}$ (Eq. \ref{eq:tau_w}). This is shown in Fig.$\:$\ref{fig:b_q2} 
where 
we plot the dependence of $b_{\phi}$ on $Q^2$ 
using Eq.$\:$\ref{eq:phi_taufit} and the relation in Eq.$\:$\ref{eq:tau_w} 
at two
values of $W$. The lower curve, at $W=2.3$ GeV, corresponds to our kinematics 
and connects our 
measurements with photoproduction values. The upper curve is closer to the 
Cornell kinematics.

Because the value of $c\Delta \tau$ is smaller than the size of the nucleon, 
the scattering
may be considered to be point-like. The application of QCD-inspired
models, sensitive to the quark structure of the interacting meson and nucleon, 
should provide an 
interesting interpretation of the observed $b(\Delta \tau)$ and $b(Q^{2})$
dependencies. It has been argued that with increasing $Q^{2}$ the radius of 
the virtual vector meson 
will shrink \cite{Bauer}, and a corresponding decrease of $b$ should be 
observed.
At large enough $Q^{2}$, quark models \cite{Frankfurt,Kopeliovich} predict 
the decrease of the 
transverse dimension of the vector meson as $r_{V} \sim {r_{h} M}/ 
\sqrt{M^2 + Q^2}$.
The mass scale $M$ 
represents a typical hadronic mass scale, which might be as small as the 
vector meson mass, e.g. 
$M_\phi$ = 1020 MeV, but is likely to be large compared to the $Q^2$ values 
of this experiment. 
Even though we do not need to invoke an explicit $Q^2$ dependence to describe 
our data, 
we note that the effects of transverse size and fluctuation 
times are not easily separated, especially when fine binning is prohibitive 
due to
limited statistics.

\section{Summary}

The electroproduction of the $\phi$(1020) vector meson was measured for
$Q^{2}$ from 0.7 to 2.2 GeV$^2$, $W$ from 2.0 to 2.6 GeV, and $\Delta \tau$ 
from 1.8 to 4.0 
GeV$^{-1}$ ($c \Delta \tau$ from 0.35 to 0.79 fm). A sample of 197 
$\phi$(1020) mesons was 
accumulated for the exclusive reaction  of $e p \rightarrow e' p' \phi$ 
with the CLAS detector in Hall B at Jefferson Lab.\\
\hfil\\
\hspace*{0.0in}(i) Taken together with the world 
data sample, we find a consistent picture of $\phi$ production on the proton.
Yet the scarcity of $\phi$ data do not
permit a precise quantitative description of the production process.\\
\hfil\\
\hspace*{0.0in}(ii) We observe the expected decrease of the slope $b_{\phi}$ 
of 
$d \sigma / dt'$ on the formation length $c \Delta \tau$ below 2 fm. The rate 
of the 
$b_{\phi}$ decrease is similar to that in $\rho$ meson production, but with a 
lower asymptotic value.
Using a simple geometric model, the data show that the interaction size of 
$\phi$ mesons 
with a proton is smaller than for $\rho$ mesons.\\
\hfil\\
\hspace*{0.0in}(iii) The $\phi$ production cross section measurement adds 
new information at low values 
of $Q^{2}$ and $W$. The cross section dependence on $Q^2$ is qualitatively 
reproduced by pomeron 
exchange models. The cross section dependence on $W$ as $W^{0.2\pm0.1}$ at 
$Q^2 = 1.3$ GeV$^2$ was 
determined by comparison to $\phi$ production at HERA after correcting for 
threshold effects. 
This dependence is the same as observed in photoproduction.\\

Additional electro- and photoproduction data from CLAS are 
currently being analyzed and will increase the overall qualitative and 
quantitative
understanding of the physics that underlies vector meson production.

\acknowledgments

We would like to acknowledge the efforts of the Accelerator, the Physics
Division, and the Hall B technical staff that made this experiment possible. 
This work was 
supported in part by the U.S. Department of Energy, including DOE Contract 
\#DE-AC05-84ER40150,
the National Science Foundation, the 
French Commissariat $\grave{\rm a}$ l'Energie Atomique, the Italian Istituto 
Nazionale di 
Fisica Nucleare, and the Korea Science Engineering Foundation. We would also 
like to 
acknowledge useful discussions with D. Cassel, M. Pichowsky and 
B. Z. Kopeliovich.

\appendix
\section{Notation}

We denote the four-momenta of the incident and scattered electron by 
$p_e$ and $p_{e'}$, the virtual
photon by $q \equiv p_e - p_{e'}$, and the target and recoil proton by 
$p_p$ and $p_{p'}$.
Each four-vector can be written as ($E$, $\vec{p}$) with appropriate 
subscripts.
We use the common notation for Lorentz invariants: $Q^2 = - q^2 > 0$, 
$\nu = q \cdot p_{p}/M_p$
($M_p$ is the mass of the proton), the squared hadronic center-of-mass 
energy $W^2 = (q + p_p)^2$, and
$t = (p_p - p_{p'})^2$ is the four-momentum transfer to the target. 
The above-threshold momentum
transfer is given by $t' = t - t_{min}(Q^2,W) < 0$, where $-t_{min}$ is 
the minimum value of $-t$ for
fixed kinematics.

\begin{figure}[htb] 
\vspace{5.8cm}
\includegraphics{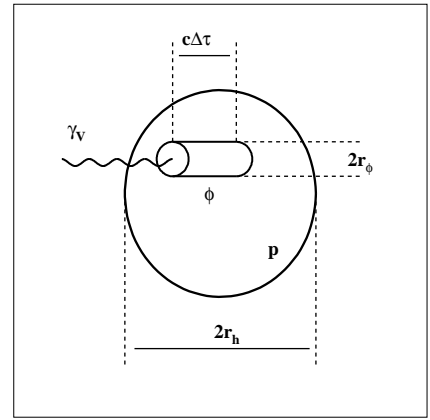}
\caption{\small{Space-time picture of the $\gamma_{V}p$ scattering through 
the conversion of 
the virtual photon into the virtual $\phi$ meson inside the target proton.}}
\label{fig:space_time}
\end{figure}

\section{Geometric Model}
\label{sec:model}

We describe a qualitative picture of vector meson diffractive scattering 
within
a simple geometric model.  A sketch of the process is shown in 
Fig.$\:$\ref{fig:space_time}. 
The virtual photon is converted into the virtual vector meson 
(of radius $r_{V}$), which diffractively 
interacts with the proton (of radius $r_{h}$) during a formation time 
$\Delta \tau$.
Differential elastic cross sections are closely related to the charge 
form factors $F(t)$ of colliding 
hadrons at high energy \cite{Goulianos,Povh}. For small values of $t$, 
the form factors are related to 
the charge radii $\langle r^2 \rangle$, via
\begin{eqnarray}
F(t) = 1 + {{1} \over {6}} \; \langle r^2 \rangle \; t + O(t^2)\: . 
\label{eq:opt_int_size}
\end{eqnarray}
For hadron-hadron elastic scattering \cite{Goulianos}, the cross sections 
depend exponentially on t: 
\begin{eqnarray}
{{d\sigma/dt} \over {(d\sigma/dt)_{t=0}}} = e^{bt} \: .
\label{eq:el_int_size}
\end{eqnarray}

Comparison of equations \ref{eq:opt_int_size} and \ref{eq:el_int_size},
and noting that the cross section is proportional
to the square of the form factor,
leads to a relationship between the radius of interaction, $R^{int}_{V}$, 
and the $t$-slope parameter $b$:
\begin{eqnarray}
b & = & { {{1} \over {3}} \left( R^{int}_{V} \right)^{2}} \; .
\label{eq:size}
\end{eqnarray}
The radius of interaction can be written as 
\begin{eqnarray}
(R^{int}_{V})^2 & \propto & \langle r_{h}^2 \rangle + \langle r_{V}^2(Q^{2}) 
\rangle \: ,
\end{eqnarray}
where $r_{h}$ and  $r_{V}$ are the radii of the nucleon and vector meson, 
respectively.

\begin{figure}[htb] 
\vspace{6.5cm}
\includegraphics{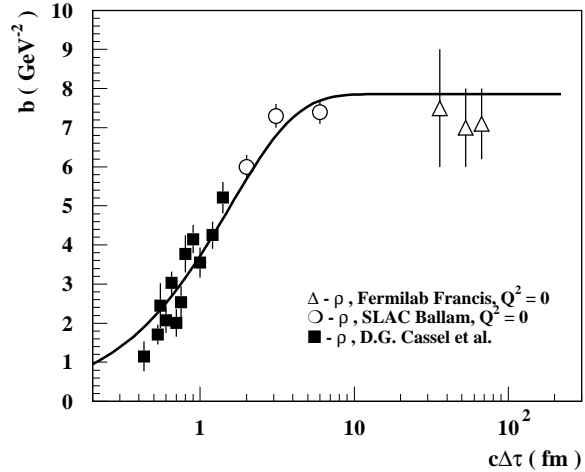}
\caption{{\small The $t$-slope parameter dependence on $c \Delta \tau$ for 
selected photo- and electroproduction
data of $\rho$ mesons. The data show a clear decrease of $b$ with decreasing 
$c \Delta \tau$ 
below 2 fm. The curve is a fit to Eq.$\:$\ref{eq:int_size}. The 
photoproduction data SLAC Ballam are from 
Ref. [38] and Fermilab Francis from Ref. [46]. The electroproduction data 
CORNELL Cassel are from Ref. [15].}}
\label{fig:b_dt_prev}
\end{figure}

Because of the virtuality of the vector meson, the interaction region 
should also decrease 
if the formation distance is less than the size of the nucleon 
(c$\Delta \tau$ 
{\tiny $\stackrel{\textstyle<}{\sim}$} $2r_{h} \approx$ 2 fm).
A representative sample of the large body of $\rho$ data shown in 
Fig.$\:$\ref{fig:b_dt_prev} 
suggests the following phenomenological 
parameterization for the $t$-slope dependence on $c \Delta \tau$:
\begin{eqnarray}
 b(c\Delta \tau) & =  & {1 \over 3} \; ( 1 - e^{- c\Delta \tau / 2 r_{h}} )
 \left( R^{int}_{V} \right)^{2} \; .
\label{eq:int_size}
\end{eqnarray}
A two-parameter fit to Eq.$\:$\ref{eq:int_size}, ignoring any explicit 
dependence of $r_{\rho}$ on $Q^2$, yields 
\begin{eqnarray} 
b_{\rho}(c \Delta \tau) = (7.86 \pm 0.26)
\left[ \; 1 - e^{-c\Delta \tau / 2 (0.78 \pm 0.05)} \; \right]
\label{eq:rho_taufit} 
\end{eqnarray}
with $\chi^{2}/DF=2.08$. 

\begin{figure}[htb] 
\vspace{6.5cm}
\includegraphics{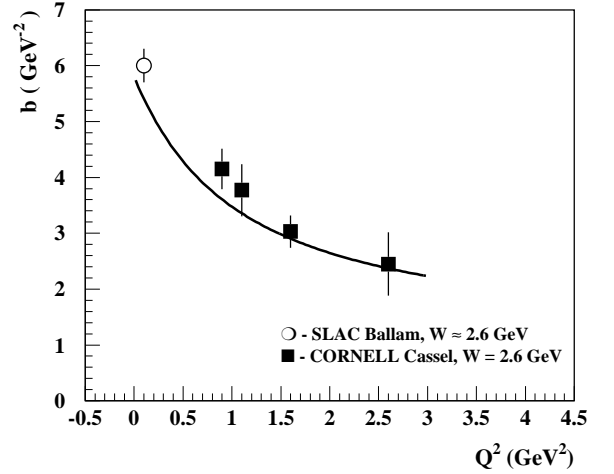}
\caption{{\small The $t$-slope parameter dependence on $Q^{2}$ for the photo- 
and electroproduction
of $\rho$ mesons at $W$ = 2.6 GeV. The data show a clear decrease of $b$ 
with increasing $Q^{2}$.
The curve is a fit to Eq.$\:$\ref{eq:int_size}. The photoproduction data 
SLAC Ballam are from 
Ref. [38]. The electroproduction data CORNELL Cassel are from Ref. [15]. }}
\label{fig:b_q2_prev}
\end{figure}

However, Eq.$\:$\ref{eq:int_size} also has an indirect dependence on $Q^2$ 
through $c \Delta \tau$. At 
fixed $W$, we can write Eq.$\:$\ref{eq:dtau} as
\begin{eqnarray}
c \Delta \tau & = & {{c \; (W^2 - M_p^2 + Q^2)} \over 
{M_p (Q^2 + M_{\rho}^2) }}
\label{eq:tau_w}
\end{eqnarray}
Thus, we can plot Eq.$\:$\ref{eq:int_size} as a function of $Q^2$, using 
this expression for $c \Delta \tau$. 
This is shown in Fig.$\:$\ref{fig:b_q2_prev} for $\rho$ data at the fixed 
value of $W$ = 2.6 GeV \cite{Cassel}. 
Thus, we see that most, if not all, of the variation of the slope parameter 
$b$ can be accounted for by changes 
in the fluctuation time.

For the kinematics of this experiment, c$\Delta \tau$ ($\approx 0.5$ fm) is 
small compared to 
the size of the nucleon, so we expect the fluctuation time factor to be 
significant for our 
$\phi$ data.

\end{document}